\documentclass[twocolumn]{revtex4}
\usepackage{graphicx,amsmath,amssymb,mathrsfs}
\usepackage{epsfig}

\newcommand{\mR}{\mathbb{R}}
\renewcommand{\eqref}[1]{(\ref{eq:#1})}

\begin{document}

\title{Extreme event statistics of daily rainfall: Dynamical systems approach}

\author{G.\ Cigdem Yalcin}
\affiliation{Department of Physics, Istanbul University, 34134, Vezneciler,
Istanbul, Turkey}

\author{Pau Rabassa and Christian Beck}
\affiliation{Queen Mary University of London, School of Mathematical Sciences, Mile End Road, London E1 4NS, UK}

\begin{abstract}
We analyse the probability densities of daily rainfall amounts at
a variety of locations on the Earth. The observed distributions of the amount of rainfall
fit well to a $q$-exponential distribution with exponent
$q$ close to $q\approx 1.3$.
We discuss possible reasons for the emergence
of this
power law.
On the contrary,
the waiting time distribution between rainy days is observed
to follow a near-exponential
distribution. A careful investigation
shows that a $q$-exponential with $q\approx 1.05$
yields actually the best fit of the data.
A Poisson process where the rate fluctuates slightly in a
superstatistical way is discussed as a possible model for this.
We discuss the
extreme value statistics for extreme daily rainfall, which
can potentially lead to flooding. This is described by
Fr\'echet distributions as the corresponding distributions of the amount of daily rainfall
 decay with a power law. On the other hand, looking at extreme event statistics of waiting times between
rainy days (leading to droughts for very long dry periods) we obtain from the observed near-exponential decay
of waiting times an extreme event statistics close to Gumbel distributions. We discuss
superstatistical dynamical systems as simple models in this context.
\end{abstract}

\maketitle

\section{Introduction}

The statistical analysis of precipitation data is an interesting problem
of major environmental importance \cite{masutani,higgins1,higgins2,silva}.
Of particular interest are extreme events of rainfall, which can
lead to floodings if a given threshold is exceeded.
From a mathematical and statistical point of view, it is natural
to apply extreme value statistics to measured rainfall data, but it is not so
clear which class of the known extreme value statistics, if any, is applicable,
and how the results differ from one spatial location to another.
Another interesting quantity to look at is the waiting time between rainy days.
Extreme events of waiting times in this context correspond to droughts.
So an interesting question is what type of drought statistics is implied by
the observed distribution of waiting times between rainfall periods if this is extrapolated to very long
waiting times.

In this paper we present a systematic investigation of the
probability distributions of the amount of daily rainfall
at variety of different locations on the Earth, and of waiting time
distributions on a scale of days and hours. Our main result is
that the observed distributions of the amount of rainfall are well-fitted by
$q$-exponentials, rather than exponentials,
which suggests that techniques
borrowed from nonextensive statistical mechanics \cite{tsallis-book} and
superstatistics \cite{beck-cohen}
could be potentially useful to better understand the rainfall statistics.
An entropic exponent of $q \approx 1.3$ is typically observed. In fact,
based on the fact that $q$-exponentials asymptotically decay with
a power law, we discuss the corresponding extreme value statistics
that is highly relevant in the context of floodings produced by extreme
rainfall events. We also investigate the waiting
time distribution between rainy events, which is much better
described by an exponential,
although an entropic exponent close
to 1 such as $q \approx 1.05$ seems to give the best fits.
We discuss possible dynamical reasons for the occurrence of $q$-exponentials in this context.

One possible reason could be superstatistical fluctuations
of a variance parameter or a rate parameter. Let us explain this a bit further.
Superstatistical techniques have been discussed in many papers
\cite{beck-cohen, swinney, touchette, jizba, chavanis, frank, celia,
straeten, mark, hanel, guo, souza, tsallis1, kaniadakis}
and they represent a powerful method to model and/or analyse complex systems
with two (or more) clearly separated time scales in the dynamics.
The basic idea is to consider for the theoretical modelling a superposition
of many systems  of statistical mechanics in local equilibrium, each with its own
local variance parameter $\beta$, and finally perform an average over the
fluctuating $\beta$. The probability density of $\beta$ is denoted by
$f(\beta)$. Most generally, the parameter $\beta$ can be any system parameter
that exhibits large-scale fluctuations, such as energy dissipation in a turbulent
flow, or volatility in financial markets. Another possibility is to
regard $\beta$ as the rate parameter of a local Poisson process,
as done, for example, in \cite{briggs}. Ultimately all expectation
values relevant for the complex system under consideration are
averaged over the distribution $f(\beta)$.
Many applications have been described in the past, including modelling
the statistics of classical turbulent flow \cite{prl2001, prl2007,
reynolds, swinney}, quantum turbulence \cite{miah}, space-time granularity
\cite{jizba2}, stock price changes \cite{straeten}, wind velocity
fluctuations \cite{rapisarda}, sea level fluctuations \cite{pau1},
infection pathways of a virus \cite{itto}, migration trajectories
of tumour cells \cite{metzner}, and much more \cite{chavanis,
abul-magd, soby, briggs, chen, cosmic}.
Superstatistical systems, when integrated over the fluctuating parameter,
are effectively described by more general versions of
statistical mechanics, where formally the Boltzmann-Gibbs entropy
is replaced by more general entropy measures \cite{hanel,souza}.
The concept can also be generalized to general dynamical systems with
slowly varying system parameters, see \cite{penrose} for some recent rigorous results in
this direction.

Our main goal in this paper is to better understand the extreme event statistics of
rainfall at various example locations on the Earth. We will start with a careful
analysis of experimentally recorded time series of the amount of rainfall measured
at a given location, whose probability density is highly relevant to model the corresponding
extreme event statistics \cite{LLR89, EKM97, Coles01, HF06, kantz}.
Ultimately of course all this rainfall dynamics can be formally
regarded as being produced by a highly nonlinear and high-dimensional
deterministic dynamical system in a chaotic state, producing the occasional rainfall event, hence it is useful to keep in mind
the basic results of extreme event statistics for weakly correlated events
as generated by mixing dynamical systems.
Recently
there has been much activity on the rigorous application of extreme values
theory to deterministic dynamical systems \cite{LFTV, FLTV, FF, FFT, FFT2,
HNT, HVR, GHN, Keller} and also to stochastically perturbed ones
\cite{AFV,FFTV,FV1}.
A remarkable feature of the dynamical system approach is that there
exist some correlations between events, and hence the extreme value
theory used to tackle it must account for this correlation going
beyond a theory that is just based on sequences of events that
are statistically independent. In the superstatistics approach,
correlations are also present, due to the fact that parameter changes
take place on long time scales, but the relaxation time of the system is short
as compared to the time scale of these parameter changes,
so that local equilibrium is quickly reached.

What is worked out in this paper is a comparison with
experimentally measured rainfall data, to decide which
extreme event statistics should be most plausibly applied to
various questions related to amount of rainfall and waiting times
between rainfall events.
Extreme value theory quite generally tells us
(under suitable asymptotic independence assumptions) that there are only three
possible limit distributions, namely, the Gumbel,
Fr\'echet and Weibull  distribution. But are these assumptions
of near-independence satisfied for rainfall data, and if yes, which of the above
three classes are relevant? This is the subject of this paper.
We will also discuss simple deterministic dynamical system models
that generate superstatistical processes in this context.

The paper is organized as follows.
 In Section II we present histograms of rainfall statistics, extracted
 from experimentally measured time series of rainfall at various locations on the Earth.
 What is seen is that the probability density of amount of rainfall is very well fitted by
 $q$-exponentials. We discuss the generalized statistical mechanics foundations of this based on nonextensive
 statistical mechanics with entropic index $q$, with $q \approx 1.3$.
 In section III we look at waiting time distributions (on a daily and hourly scale) between
 rainy episodes. These are observed to be close to exponential functions, similar as
 for the Poisson process. However, a careful analysis shows that a slightly better fit is
 again given by $q$-exponentials, but this time with $q$ much closer to 1.
 A simple superstatistical model for this is discussed in section IV, a Poisson process that has
 a rate parameter that fluctuates in a superstatistical way. We review standard extreme event statistics in
 section V and then, in section VI,
 based on the measured experimental results of rainfall statistics, we develop the corresponding extreme value statistics.
 In section VII
 we analyse the ambiguities that arise for the extreme event statistics of waiting times, depending on whether
 we assume the waiting time distribution is either an exact exponential or a slightly deformed
 $q$-exponential as produced by superstatistical fluctuations.
Finally, in section VIII we describe a dynamical systems approach to superstatistics.

\section{Daily rainfall amount distributions at various locations}

We performed a systematic investigation of time series of rainfall data for 8 different example locations on the Earth (Figs.~1-8).
The data are from various publicly available web sites. When doing a histogram of the amount of daily rainfall
observed, a surprising feature arises. All distributions are power law rather than exponential.
They are well fitted by so-called $q$-exponentials, functions of the form
\begin{equation}
e_q(x)=(1+(1-q)x)^{1/(1-q)}
\end{equation}
which are well-motivated by generalized versions of statistical mechanics relevant for systems with long-range interactions, temperature fluctuations and
multifractal phase space structure
\cite{tsallis-book, beck-cohen}. Of course the ordinary exponential is recovered for $q \to 1$.
Whereas the data of most locations are well fitted by $q \approx 1.3$, Central England and Vancouver have somewhat lower values of
$q$ closer to 1.13.

One may speculate what the reason for this power law is. Nevertheless, the formalism of nonextensive
statistical mechanics \cite{tsallis-book}  is designed to describe complex systems with spatial or temporal long-range interactions, and
$q$-exponentials occur in this formalism as generalized canonical distributions that maximize $q$-entropy
\begin{equation}
S_q= \frac{1}{q-1} \sum_i (1-p_i^q),
\end{equation}
where the $p_i$ are the probabilities of the microstates $i$. Ordinary statistical mechanics
is recovered in the limit $q \to 1$, where the $q$-entropy $S_q$ reduces to the
Shannon entropy
\begin{equation}
S_1 = - \sum_i  p_i \log p_i.
\end{equation}
The generalized canonical distributions maximize the $q$-entropy subject
to suitable constraints. In our case the constraint is given by the average amount of daily rainfall at a given location.
The way rainfall is produced is indeed influenced by highly complex weather systems and condensation processes in clouds,
so one may speculate that more general versions of statistical mechanics could be relevant
as an effective description. Also for
hydrodynamic turbulent systems \cite{prl2001,swinney} and pattern forming systems \cite{daniels} these generalized statistical mechanics methods
have previously been shown to yield a good effective description.
The amount of rain falling on a given day is a complicated spatio-temporal stochastic process
with intrinsic correlations, as rainy weather often has a tendency to persist for a while,
both spatially and temporally.
The actual value of $q$ for the observed rainfall statistics reflects characteristic effective properties in the climate and temporal precipitation pattern at the given location.
For temperature distributions at the same locations as in Fig.1-8, see \cite{yalcin}.

\begin{figure}
\includegraphics[scale=0.5]{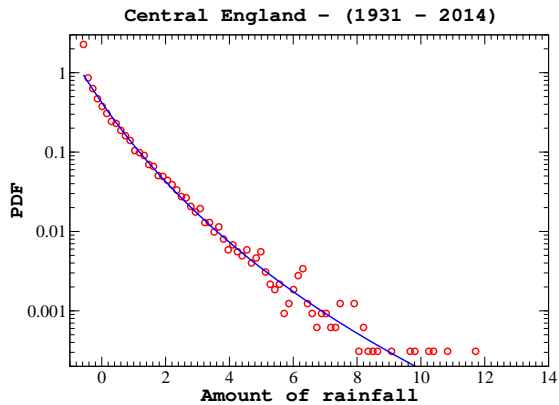}
\caption{Observed probability distribution of amount of daily rainfall (minus average value) in Central England (red data points).
The blue curve is a fit with a $q$-exponential $e_q(-x)$. The parameter $q$ is fitted for the various locations. For Central England, $q=1.13$. }
\end{figure}

\begin{figure}
$\,$

\vspace*{1cm}

\includegraphics[scale =0.5]{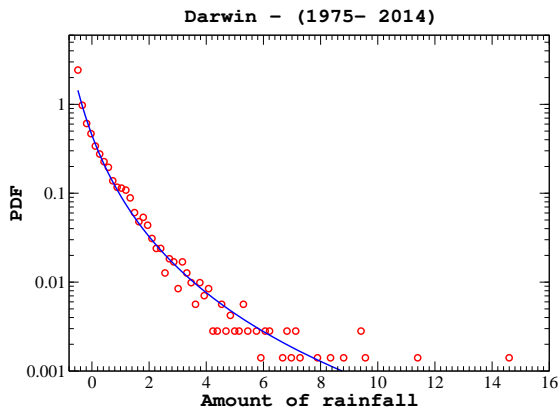}
\caption{As Fig.~1, but for Darwin ($q=1.30$).}
\end{figure}
\begin{figure}
\includegraphics[scale =0.5]{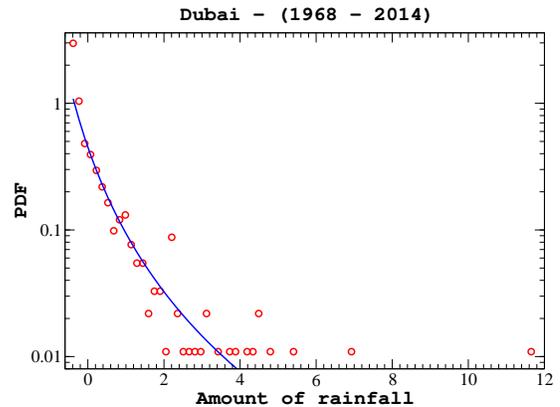}
\caption{As Fig.~1, but for Dubai ($q=1.30$).}
\end{figure}
\begin{figure}
\includegraphics[scale =0.5]{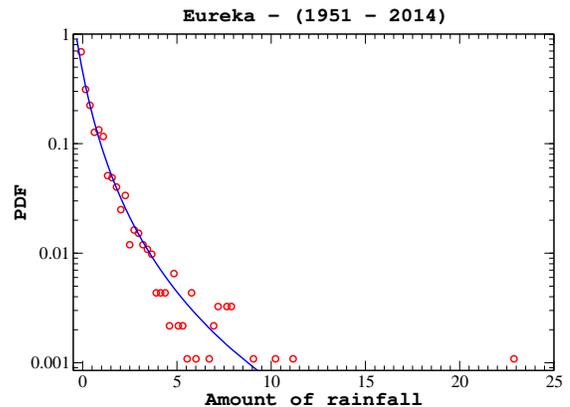}
\caption{As Fig.~1, but for Eureka ($q=1.30$).}
\end{figure}
\begin{figure}
\includegraphics[scale =0.5]{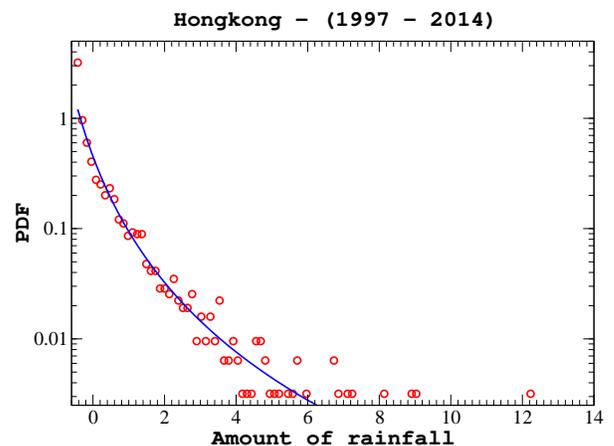}
\caption{As Fig.~1, but for Hongkong ($q=1.30$).}
\end{figure}
\begin{figure}
\includegraphics[scale =0.5]{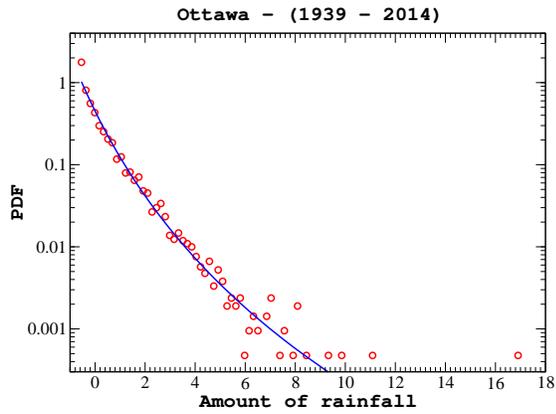}
\caption{As Fig.~1, but for Ottawa ($q=1.15$).}
\end{figure}.
\begin{figure}
\includegraphics[scale =0.5]{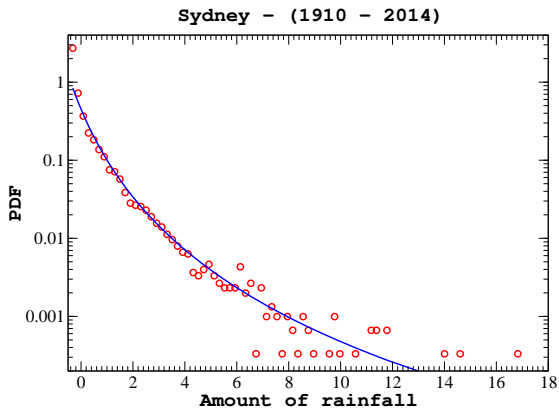}
\caption{As Fig.~1, but for Sydney ($q=1.25$).}
\end{figure}
\begin{figure}
\includegraphics[scale =0.5]{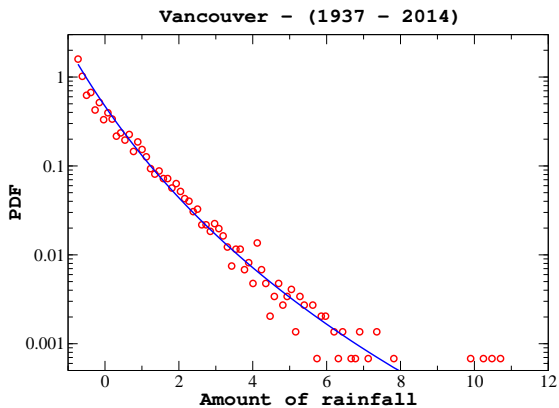}
\caption{As Fig.~1, but for Vancouver ($q=1.13$).}
\end{figure}

\section{Waiting time distributions between rainy days}
\label{sec:waiting-time}

Another interesting observable that we extracted from the data is the waiting time distribution
between rainy episodes. We did this both for a time scale of days and a time scale of hours.
A given day is marked as rainy if it rains for some time during that day. The waiting
time is then the number of days one has to wait until it rains again, this is a random variable
with a given distribution which we can extract from the data. Results
for the waiting time distributions are shown in Fig.~9-14. What one observes here is that
the distribution is nearly exponential. That means the Poisson process of nearly independent
point events of rainy days is a reasonably good model.

At closer inspection, however, one sees that again a slightly deformed $q$-exponential, this time with $q \approx 1.05$,
is a better fit of the waiting time distribution. As worked out in the next section, one may explain this with a superstatistical
Poisson process, i.e. a Poisson process whose rate parameter -- on a long time scale-- exhibits fluctuations
that are $\chi^2$-distributed, with a rather large number of degrees of freedom $n$.
\begin{figure}
\includegraphics[scale =0.5]{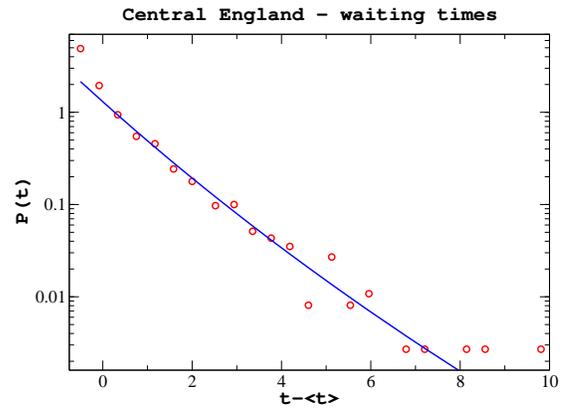}
\caption{Waiting time distribution between rainy days for Central England (red data points). The blue curve is a fit
with a $q$-exponential. For waiting time distributions, the parameter $q$ is much closer to 1 ($q=1.05$ for Central England).}
\end{figure}
\begin{figure}
\includegraphics[scale =0.5]{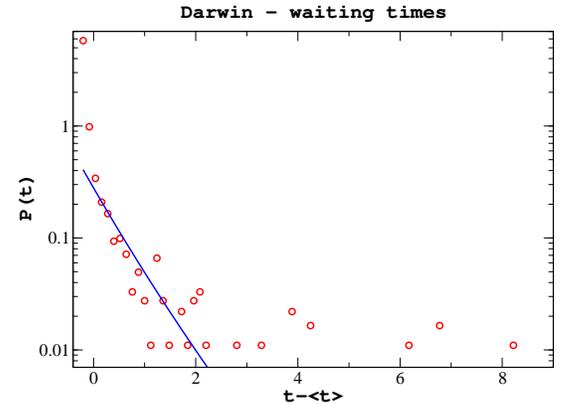}
\caption{As Fig.~9, but for Darwin ($q=1.05$).}
\end{figure}
\begin{figure}
\includegraphics[scale =0.5]{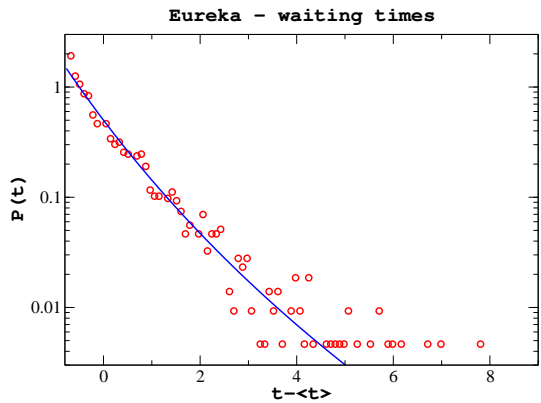}
\caption{As Fig.~9, but for Eureka ($q=1.10$).}
\end{figure}
\begin{figure}
\includegraphics[scale =0.5]{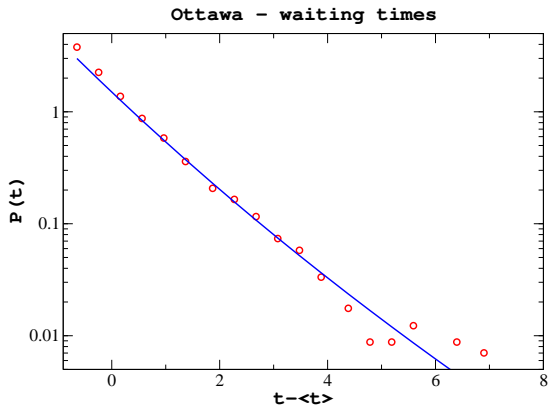}
\caption{As Fig.~9, but for Ottawa ($q=1.05$).}
\end{figure}

\vspace*{0.8cm}

\begin{figure}
\includegraphics[scale =0.5]{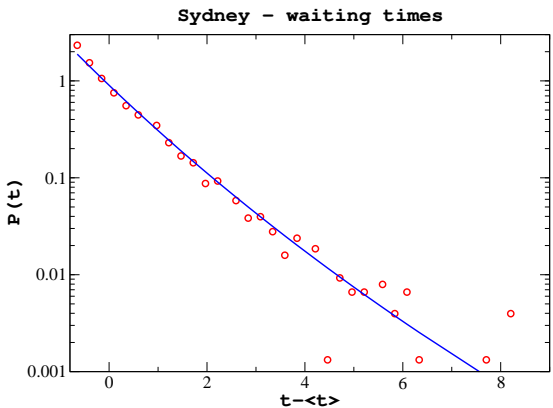}
\caption{As Fig.~9, but for Sydney ($q=1.06$).}
\end{figure}
\begin{figure}
\includegraphics[scale =0.5]{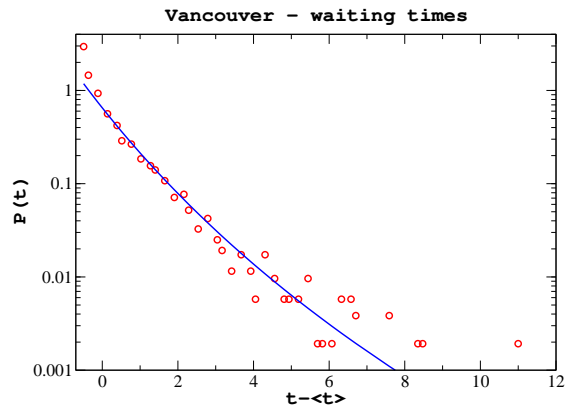}
\caption{As Fig.~9, but for Vancouver ($q=1.10$).}
\end{figure}

\section{Superstatistical Poisson process}

We start with a very simple model for the return time of rainfall events
(or extreme rainfall events) on any given time scale. This is
to assume that the events follows a Poisson process.
For a Poisson process the waiting times are exponentially
distributed,

\[ p(t | \beta) = \beta \exp(-\beta t).\]

Here, $t$ is the time from one event (peak over threshold) to the
next one, and $\beta$ is a positive parameter,
the rate of the Poisson process. The symbol $p(t|\beta)$ denotes the
conditional probability density to observe a return time $t$ provided the parameter
$\beta$ has certain given value.

The key idea of the superstatistics approach \cite{beck-cohen, briggs}
can be applied to this simple model, thus constructing a superstatistical
 Poisson process. In this case the parameter $\beta$ is regarded as a fluctuating random variable as well, but these fluctuations take place on a very large time scale. For example,
 for our rainfall statistics the time scale on which $\beta$ fluctuates may correspond to weeks (different weather
 conditions) whereas our data base records rainfall events on an hourly basis.

If $\beta$ is distributed with probability density $f(\beta)$, and
fluctuates on a large time scale, then one obtains the marginal distribution
of the return time statistics as
\begin{equation}
\label{eq:marg-ss-dist}
p(t) = \int_0^\infty f(\beta) p(t | \beta) =  \int_0^\infty f(\beta)
\beta \exp(-\beta t).
\end{equation}

This marginal distribution is actually what is recorded
when we sample histograms of the observational data.

By inferring directly on a simple model for the distribution $f(\beta)$,
a more complex model for the return times can be derived without much technical
complexity. For example, consider that there are $n$ different Gaussian
random variables $X_i, i=1,\dots,n$, that influence the dynamics of the intensity
parameter $\beta$ as a random variable. We may thus assume as a very simple model
that $\beta=\sum_{i=1}^n X_i^2$ with $E(X_i)=0$ and $E(X_i^2) \neq 0$. Then the
probability density of $\beta$ is given by a $\chi^2$-distribution:
\[
f(\beta) = \frac{1}{\Gamma(n/2)} \left(\frac{n}{2\beta_0} \right)^{n/2}
\beta^{n/2-1} \exp\left(\frac{n\beta}{2\beta_0} \right),
\]
where $n$ is the number of degrees of freedom and $\beta_0$ is a shape parameter
that has the physical meaning of being the average of $\beta$ formed with the distribution $f(\beta)$.

The integral \eqref{marg-ss-dist} is easily evaluated and one obtains the
$q$-exponential distribution:
\[
p(t) \sim (1+b(q-1) t)^{1/(1-q)},
\]
where $q=1+2/(n+2)$ and $b= 2 \beta_0/(2-q)$.

To sum up, this model generates $q$-exponential distributions by a simple
mechanism, fluctuations of a rate parameter $\beta$.
Typical $q$-values obtained in our fits are $q=1.3$ for rainfall
amount and $q= 1.05$ for waiting time between rainfall events.

\section{Extreme value theory for stationary processes}
\label{sec:EVT}

Classic extreme value theory is concerned with the probability
distribution of unlikely events. Given a stationary stochastic
process $X_1,X_2, \dots$, consider the random variable
$M_n$ defined as the maximum over the first $n$-observations:

\begin{equation}
\label{eq:max process}
M_n = \max(X_1,\dots,X_n).
\end{equation}

In many cases the limit of the random variable $M_n$ may
degenerate when $n\rightarrow \infty$. Analogously to central
limit laws for partial sums, the degeneracy of the limit can be
avoided by considering a rescaled sequence $a_n(M_n - b_n)$ for
suitable normalising values $a_n\geq 0$ and $b_n\in \mR$. Indeed,
extreme value theory studies the existence of normalising values such that

\begin{equation}
\label{eq:limit-EV}
 P\left(a_n\left(M_n-b_n\right) \leq x \right) \rightarrow G\left(x\right).
\end{equation}
as $n\rightarrow \infty$, with $G(x)$ a non-degenerate probability distribution.

Two cornerstones in Extreme Value Theory are the Fisher-Tippet Theorem \cite{fisher}
and the Gnedenko Theorem \cite{gnedenko}. The former asserts that if the limiting
distribution $G$ exist, then it must be either
one of three possible types, whereas the latter theorem gives necessary
and sufficient conditions for the convergence of  each of the types.
A third cornerstone in Extreme Value Theory are the Leadbetter
conditions \cite{Leadbetter74, LLR89}.
These are a kind of weak asymptotic independence conditions,
under which the two previous theorems generalize to stationary stochastic
series satisfying them. Let us review these results in somewhat more detail.


In the case where the process $X_i$ is independent identically
distributed (i.i.d.) the Fisher-Tippett Theorem states that
if $X_1, X_2, \dots$ is i.i.d. and there exist sequences
$a_n\geq 0$ and $b_n\in \mR$ such that the limit distribution
$G$ is non-degenerate, then it belongs to one of the following types:
\begin{description}
\item[Type I  :] $G(x) = \exp\left(- e^{-x} \right)$ for $x\in\mR$. This distribution is known as the
{\it Gumbel} extreme value distribution (e.v.d.).

\item[Type II :] $G(x) = \exp\left(-x^{-\alpha}\right)$, for $x>0$;
$G(x)=0$, otherwise; where $\alpha >0$ is a parameter. This family of distributions is known
as the {\it Fr\'echet} e.v.d.

\item[Type III: ]
$ G(x) = \exp \left(-(-x)^\alpha\right)$, for $x\leq0$; $G(x)=1$, otherwise; where
$\alpha >0$ is a parameter. This family is known as the {\it Weibull} e.v.d.
\end{description}

A further extension of this result is the Gnedenko Theorem, which provides a
characterization of the convergence in each of these cases.  Let $X_1,X_2,\dots$ be an i.i.d.
stochastic process and let $F$ be its cumulative distribution
function. Consider $x_M=\sup\{x|\, F(x) < 1\}$.  The following conditions
are necessary and sufficient for the convergence to each type of e.v.d.:
\begin{description}
\item[Type I:]
There exists some strictly positive function $h(t)$ such that
$\lim_{t \rightarrow x_M^-} \frac{1-F(t+xh(t))}{1-F(t)} = e^{-x}$
for
 all real $x$;
\item[Type II:]
$x_M= +\infty$ and $\lim_{t\rightarrow \infty} \frac{1-F(tx)}{1-F(t)}= x^{-\alpha}$, with
$\alpha>0$, for each $x>0$;
\item[Type III:]
$x_M < \infty$ and $\lim_{t\rightarrow 0} \frac{1-F(x_M - tx)}{1-F(x_M - t)}= x^{\alpha}$, with
$\alpha>0$, for each $x>0$.
\end{description}

This result implies that the extremal type is completely determined
by the tail behaviour of the distribution $F(x)$.

\section{Extreme event statistics for exponential and q-exponential distributions}

The rainfall data were well described by $q$-exponentials, but waiting time
distributions were observed to be close to ordinary exponentials, with $q$ only
deviating by a small amount from 1. Let us now discuss
the differences in extreme value statistics that arise from theses different
distributions.

In the case where $X$ is distributed as an ordinary exponential
function with parameter $\lambda$, we have
\begin{equation}
\label{eq:hazard-exp}
1-F(t) = \left\{
  \begin{array}{rcl}
    \exp(-\lambda t) & \text{ if } & t>0 \\
                   1 & \text{ if } & t<0.
  \end{array}
\right.
\end{equation}
It is not difficult to check that the exponential distribution belongs to
the Gumbel domain of attraction. In other words, the extreme events associated
to the exponential distribution will be Gumbel distributed.

Recall that the $q$-exponential function is defined as
\[
\exp_q (t) :=\left[1+(1-q) t \right]^{1/(1-q)},
\]
with $1\leq q < 2$.
A random variable $X$ is $q$-exponential distributed (with parameter $\lambda$)
if its density function is equal to $(2-q)\lambda \exp_q(-\lambda x)$.
In such a case, its hazard function is
\begin{equation}
\label{eq:hazard-q-exp}
1-F(t) = \left\{
  \begin{array}{rcl}
     \exp_{q'} \left(-\lambda t /q'\right)  & \text{ if } & t>0 \\
                                          1 & \text{ if } & t<0.
  \end{array}
\right.
\end{equation}
where $q' = 1/(2-q)$.

Using the Gnedenko theorem if follows that the $q$-exponential distribution
belongs to the Fr\'echet domain of attraction. In this case the shape
parameter of the Fr\'echet distribution $\alpha$ is equal to
$\frac{2-q}{q-1}$.

\section{Model uncertainty}

Extremely large waiting times for rainfall events correspond to droughts.
Clearly, it is interesting to extrapolate our observed waiting time distributions
to very large time scales.
However, in Section~\ref{sec:waiting-time} we saw that in most cases it is difficult to
discern if the waiting time distribution is that of a Poisson process, distributed
as an exponential, or if it is a $q$-exponential with $q$ close to one.
This can make a huge difference for extreme value statistics.
The aim of this section is to assess the impact of choosing one or
the other model.

Consider $k$ a constant and $X$ a random variable modelled either
by an exponential or a $q$-exponential. To normalize the problem,
we can scale our analysis in terms of the mean. In other words,
we look at the probability of $X$ being bigger than a multiple, say $k$-times, the
mean of $X$.

If $X$ is distributed like an exponential with parameter $\lambda$,
its mean is equal to $1/\lambda$ and its hazard function is given by
\eqref{hazard-exp}. Then it is easy to check that
\[
P(X> k E(X)) = \exp\left(-k\right).
\]

On the other hand, if $X$ is distributed like a $q$-exponential
with parameter $\lambda$, its mean is equal to $1/\lambda(3-2q)$
(provided $1\leq q < 3/2$)
and its hazard function is given by \eqref{hazard-q-exp}.
In this case we have
\[
P(X> k E(X)) = \exp_{q'}\left(-\frac{2-q}{3-2q} k\right).
\]

\begin{figure}
\label{fig:returns}
\begin{center}
\includegraphics[scale =0.7]{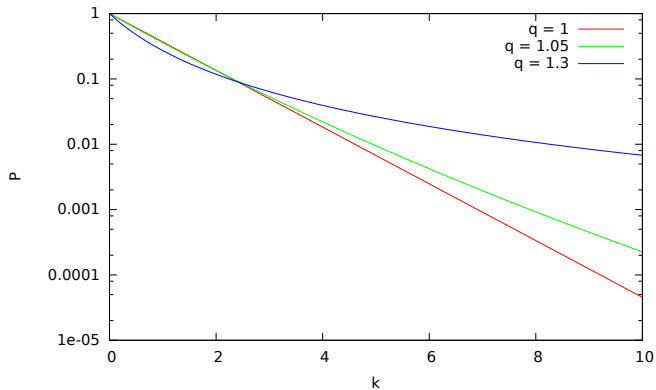}
\end{center}
\caption{Horizontal axis: Multiple $k$ of the mean,
vertical axis: probability $P$ for an event at that level.}
\end{figure}

Recall that the exponential distribution can be understood as the
limit of the $q$-exponential as $q$ goes to $1$. This is also
true for the probability above, which converges to $\exp\left(-k\right)$
as $q$ goes to $1$.
In Fig 15 we plotted the probability of an event
of level $k$ for different values of $q$.
For instance the probability of having
an observation bigger than $5$ times the mean is $0.0068$ for $q=1$, $0.0095$ when
$q=1.05$ and $0.026$ when $q=1.3$. When we look at the probability of an
observation bigger than $10$ the mean, it is $0.000045$, $0.00023$ and
$0.0068$ respectively. Apparently, the predicted drought statistics is very different
choosing either the value $q=1$ or the very similar value $q=1.05$
for the observed waiting time distribution. This illustrates the general uncertainty
in model building for extreme rainfall and drought events \cite{pau2}.

\section{Dynamical systems approach}

Ultimately, the weather and rainfall events at a given location can be regarded as
being produced by a very high-dimensional deterministic dynamical system exhibiting chaotic
properties. It is therefore useful to extend the superstatistics concept
to general dynamical systems, following similar lines
of arguments as in the recent paper \cite{penrose}.

The basic idea here is that one has a given dynamics (which, for simplicity, we
take to be a discrete mapping $f_a:\Omega \to \Omega$ on some phase space $\Omega$) which
depends on some control parameter $a$. If $a$ is changing on a large time scale,
much larger than the local relaxation time determined by the Perron-Frobenius operator of the mapping,
then this dynamical system with slowly changing control
parameter will ultimately generate a superposition of invariant densities for the given
parameter $a$. Similarly, if we can calculate return times to certain particular regions of the
phase space for a given parameter $a$, then in the long term the return time distribution
will have to be formed by taking an average over the slowly varying parameter $a$.
Clearly, the connection to the previous sections is that a rainfall event
corresponds to the trajectory of the dynamical system being in a particular subregion of the phase space $\Omega$,
and the control parameter $a$ corresponds to the parameter $\beta$ used in the previous sections.

Let us consider families of maps $f_a$ depending on a control parameter
$a$. These can be {\em a priori} arbitrary maps in arbitrary dimensions, but it is useful to
restrict the analysis to mixing maps and assume that an absolutely continuous invariant density
$\rho_a(x)$ exists for each value of the control parameter $a$.
The local dynamics is
\begin{equation}
x_{n+1}=f_a(x_n).
\end{equation}
We allow for a time dependence of $a$ and study the long-term
behavior of iterates given by
\begin{equation}
x_n=f_{a_n}\circ f_{a_{n-1}} \circ \ldots f_{a_1} (x_0). \label{it}
\end{equation}
Clearly, the problem now requires the specification of the
sequence of control parameters $a_1, \ldots , a_n$ as well,
at least in a statistical sense. One possibility is a periodic
orbit of control parameters of length $L$. Another possibility
is to regard the $a_j$ as random variables and to specify the
properties of the corresponding stochastic process in parameter space.
This then leads to a distribution of parameters $a$.

In general, rapidly fluctuating parameters $a_j$ will lead to a very
complicated dynamics. However, there is a significant simplification
if the parameters $a_j$ change slowly. This is the analogue of
the slowly varying temperature parameters in the superstatistical
treatment of
nonequilibrium statistical mechanics \cite{beck-cohen}.
The basic assumption of superstatistics is that an environmental control
parameter $a$ changes only very slowly, much slower than
the local relaxation time of the dynamics. For maps this means that significant changes
of $a$ occur only over a large number $T$ of iterations.
For practical purposes one can model this superstatistical case as follows:
One keeps $a_1$ constant for $T$ iterations ($T>>1$), then switches
after $T$ iterations to a new value $a_2$, after $T$ iterations
one switches to the next values $a_3$, and so on.

One of the simplest examples is a period-2
orbit in the parameter space. That is, we have an alternating sequence
$a_1,a_2$ that repeats itself, with switching between the two possible values
taking place after $T$ iterations.
This case was given particular attention in \cite{penrose}, and
rigorous results were derived for special types of maps where the invariant
density $\rho_a$ as a function of the parameter $a$ is under full control,
so-called Blaschke products.

Here we discuss two important examples, which are of importance in the context of the
current paper, namely how to generate (in a suitable limit) a superstatistical
Langevin process, as well as a superstatistical Poisson process, using
strongly mixing maps.

{\bf Example 1} {\bf Superstatistical Langevin-like process} We take for $f_a$ a map of
linear Langevin type \cite{roep,dyna}.
This means $f_a$ is a 2-dimensional map given by a skew product of the form
\begin{eqnarray}
x_{n+1}&=&g(x_n) \label{ggg} \\
y_{n+1}&=&e^{-a\tau}y_{n}+\tau^{1/2}(x_n-\bar{g})
\end{eqnarray}
Here $\bar{g}$ denote the average of iterates of $g$.
It has been shown in \cite{roep} that for $\tau \to 0$, $t=n\tau$ finite
this deterministic chaotic map generates a dynamics equivalent to a linear
Langevin equation \cite{vKa}, provided the map $g$ has
the so-called $\varphi$-mixing property \cite{billingsley}, and regarding the initial
values $x_0\in[0,1]$ as a smoothly
distributed random variable. Consequently, in this limit the
variable $y_n$ converges to the Ornstein-Uhlenbeck
process \cite{roep,dyna}
and its stationary density is given by
\begin{equation}
\rho_\beta(y) =\sqrt{\frac{\beta}{2\pi}} e^{-\frac{1}{2}\beta y^2}
\end{equation}
The variance parameter $\beta$ of this Gaussian depends on the map $g$ and
the damping constant $a$.
If the parameter $a$
changes on a very large time scale, much larger than
the local relaxation time to equilibrium, one expects
for the long-term distribution of iterates a
mixture of Gaussian distributions with different variances $\beta^{-1}$.
For example, a period 2 orbit of parameter changes yields
a mixture of two Gaussians
\begin{equation}
p(y)=\frac{1}{2}\left( \sqrt{\frac{\beta_1}{2\pi}} e^{-\frac{1}{2}\beta_1 y^2}+\sqrt{\frac{\beta_2}{2\pi}}
e^{-\frac{1}{2}\beta_2y^2} \right) .\label{two-gau}
\end{equation}
Generally, for more complicated parameter changes
on the long time scale $T$,
the long-term distribution
of iterates $y_n$ will be given by a mixture of Gaussians
with a suitable weight function $h (\beta)$ for $\tau \to 0$:
\begin{equation}
p(y)\sim \int d\beta \; h(\beta)e^{-\frac{1}{2}\beta y^2}
\end{equation}
This is just the usual form of superstatistics used in
statistical mechanics, based on a mixture of Gaussians with
fluctuating variance with a given weight function \cite{beck-cohen}.
Thus for this example of skew products
the superstatistics of the map $f_a$ reproduces the concept
of superstatistics in nonequilibrium statistical mechanics,
based on the Langevin equation. In fact, the map $f_a$ can be
regarded as a possible microscopic dynamics underlying the Langevin equation.
The random forces pushing the particle left and right are in this case
generated by deterministic chaotic map $g$ governing the dynamics of the variable
$x_n$. Generally it is possible to consider any
$\varphi$-mixing map here \cite{roep}.
Based on functional limit theorems, one can prove equivalence with
the Langevin equation in the limit $\tau \to 0$.

We note in passing that if the mixing property is not satisfied, then of course much more
complicated probability distributions are expected for sums of iterates of a given map $g$.
Numerical evidence has been presented that in this case often $q$-Gaussians with $q>1$ are observed \cite{tir1,tir2,tir3}.

{\bf Example 2} {\bf Superstatistical Poisson-like process}

Take $f_a:\Omega \to \Omega$ to be a strongly mixing map, for example
the binary shift map $f_a(x)=2x \mod 1$ on the interval $\Omega= [0,1]$.
Consider a very small subset of the phase space of size $\epsilon$, say $I_\epsilon= [1-\epsilon, 1]$
and a generic trajectory of the binary shift map for a generic initial value $x_0=\sum_{i=1}^\infty \sigma_i2^{-i}$, where $\sigma_i \in \{0,1\}$
are the digits of the binary expansion of the initial value $x_0$. We define a `rainfall' event to happen if this trajectory
enters $I_\epsilon$ (of course, for true rainfall events the dynamical system is much more complicated and lives on a
much higher-dimensional phase space). It is obvious that the above sequence of events follows Poisson-like statistics, as the iterates
of the binary shift map are strongly mixing, which means asymptotic statistical independence for a large
number of iterations. Indeed between successive visits of the very small interval $I_\epsilon$,
there is a large number of iterations and hence near-independence. Hence the binary shift map generates a very good approximation of the Poisson process for small enough $\epsilon$,
and the waiting time distribution between events is exponential.

We may of course also look at a more complicated system $f_a$, where we iterate a strongly mixing map $f_a$ which depends
on a parameter $a$, and where the
invariant density $\rho_a$ of the map $f_a$ depends on the control parameter in a nontrivial way. Examples are Blaschke products,
studied in detail in \cite{penrose}. If the parameter $a$ varies on a large time scale, so does the probability
$p_\epsilon=\int_{I_\epsilon} \rho_a(x)dx$ of iterates to enter the region $I_\epsilon$, and hence the rate parameter
of the above Poisson-like process will also vary. The result is a superstatistical Poisson-like process, generated
by a family of deterministic chaotic mappings $f_a$. In this way we can build up a formal
mathematical framework to dynamically generate superstatistical Poisson processes.

\section{Conclusion}

We started this paper with experimental observations:
 The probability densities of daily rainfall amounts at
a variety of locations on the Earth are not Gaussian or exponentially
distributed, but follow
an asymptotic power law, the $q$-exponential distribution.
The corresponding entropic exponent
$q$ is close to $q\approx 1.3$.
The waiting time distribution between rainy episodes is observed
to be close to an exponential distribution, but again a careful analysis shows that
a $q$-exponential is a better fit, this time with
$q$ close to 1.05.
We discussed the corresponding extreme value distributions,
leading to Gumbel and Fr\'echet distributions.
We made contact with a very important concept that is borrowed from
nonequilibrium statistical mechanics, the superstatistics approach, and
pointed out how to
generalize this concept to strongly mixing mappings that can generate Langevin-like
and Poisson-like processes for which the corresponding variance or rate parameter fluctuates in
a superstatistical way. Of course rainfall is ultimately described
by a very high dimensional and complicated spatially extended dynamical
system, but simple model systems  as discussed in this paper may help.

\section*{Acknowledgement}

G.C. Yalcin was supported by the Scientific Research Projects Coordination Unit of Istanbul University with project number 49338. She gratefully acknowledges the hospitality of Queen Mary University of London, School of Mathematical Sciences, where this work was carried out. The research of P. Rabassa and C. Beck is supported by EPSRC grant number EP/K013513/1.


\end{document}